\documentstyle[11pt,newpasp,twoside,epsfig]{article}
\def\edcomment#1{\iffalse\marginpar{\raggedright\sl#1\/}\else\relax\fi}
\reversemarginpar

\begin{document}
\title{Optical identification of supersoft X-ray sources in M31 }

\author{ Marina Orio$^{1,2}$,
Roberto Casalegno$^{1}$, Chris Conselice$^{2,3}$, Jochen Greiner$^{4}$,
Petko L. Nedialkov$^{5}$, Nikolay A. Tikhonov$^{6}$}

\affil {1. Turin Astronomical Observatory, I-10025 Pino Torinese (TO), Italy}
\affil {2.Astronomy Department,  470 N. Charter Str., 53706 Madison,
WI, USA} 
\affil {3. Hubble Space Science Telescope Institute, USA} 
\affil {4. Astrophysical Institute, Posdam, FRG}
\affil{ 5. Department of Astronomy, Sofia University, Bulgaria}
\affil {6. SAO of the Russian Academy of Sciences, Nizhnij Arkhyz, Russia}

\begin{abstract}

 We report on initial progress
in a program of optical identification of supersoft
X-ray sources in M31, pursued mainly with the WIYN telescope.
We propose the identification of one supersoft X-ray
source with a variable star, which we believe to
have been a classical or recurrent nova in outburst in September 1990.
The nova remnant must have been still a supersoft X-ray sources
5 years after this observation, when it was observed with ROSAT.   
\end{abstract}

\section{Introduction}

16 supersoft X-ray sources (SuSo) were detected during the
deep ROSAT PSPC pointings of M31 (Greiner 1996, Supper et al. 1997,
Greiner 2000). Up to 18 other ROSAT sources
in the direction of M31 have been proposed to belong
 to this class (Kahabka 1999).  Most fields were subsequently observed with
the ROSAT HRI and the positions are precise within
10-15 arcsec. It is crucial to determine the precise nature
of this population and whether it is associated with
M31.  Yet, only 9 of the sources 
have been scheduled for CHANDRA observations,
which will make the optical identification almost straightforward 
in most cases (thanks to arcsecond precision).
 For all the others, we have to rely
on high quality optical data.  The optical identification and the
subsequent optical study of these sources is
extremely necessary in order to increase our knowledge and
the statistics of the extragalactic low mass X-ray binary
population. Yet, the identification poses several
challenges. We can assess that the optical
counterpart belongs to M31 only by obtaining the spectrum
(through the red shift of the spectral lines).
However, the maximum absolute magnitude of
the supersoft X-ray sources observed in the Galaxy and the
Magellanic Clouds is M$_V$=-2 (see Greiner 2000 and
references therein). Extrapolating to the distance
to M31, we expect the most luminous members of the
class of the close binary supersoft X-ray sources described
in van den Heuvel et al. (1992) or in Kahabka and van den
Heuvel (1997) to be at  V$\simeq$22. 
The spectra can be obtained mainly, even if not always, with
very large telescopes of the new generation.

The starting point for optical identification of
supersoft X-ray sources is, however, {\it optical photometry}
(e.g. Orio et al. 1994, 1997). The supersoft X-ray sources
are expected to be very ``blue'' and hot and
be identified thanks to the color indexes (U-R, B-R, B-V),
and often to the optical variability, mostly on time scales
of few hours to a day.  Among foreground sources  
even  cooling neutron stars might be optical counterparts
(if very faint and not showing binary-type variability, e.g. Walter
et al. 1997), while apparently more luminous optical counterparts might
be AM Her stars, other CV's, PG 1159 stars
and VY Scl stars (see Greiner 2000).
 The latter are particularly interesting
because they imply an exciting extension of the class of SuSo binaries
to optically bright systems.
 Sources that truly belong to the M31 population
are: exceptional planetary nebulae, symbiotic stars and novae shortly after
thermonuclear runaways, and other close binaries. The latter
belong to two different types: long period (0.5-2 days, like
CAL 83 and 87, see van den Heuvel et al. 1992) and short period ones (with
periods of $\simeq$ 4 hours, and expected to be intrinsically
less luminous) like SMC 13 or RX J0537.7-7034 (e.g. Greiner et al. 1999).
Finally, background sources are AGN, Seyfert and other active galaxies.

\section{The optical imaging program with the WIYN telescope}

  Some of us (the first four authors in this
paper) have started an imaging program
of M31 with the WIYN 3.5 ``new technology'' telescope.
Under the best conditions we have images that reach B$\simeq$24.
 Preliminary results indicate two  likely candidates that should be
studied spectroscopically. Moreover,
we discovered  a variable that we believe to
have been a nova in M31. This nova is almost certainly
the ``culprit'' of the supersoft X-ray
emission.  It was identified thanks to comparisons with a large
number of images obtained over several years by two of
 us (P.L.N. and A.N.T.) and other published images in the literature.
Up to now only 4 classical novae and one recurrent nova
have been identified with supersoft X-ray sources.
The existence of hot white dwarf after the outburst has very
important consequences for the mechanisms and evolution
of nova systems (see accompaining poster by Orio \& Parmar).

\section{Possible candidates}

We still have very preliminary results for most sources
we studied except RX J0044.0+4118.  
We  identified possible candidates in the fields of
 RX J0039.7+4030 and
RX J0046.2+4144, respectively at U$\simeq$22 and U$\simeq$20;
however these candidates appear interesting for their U-R values
and {\it not} for their variability.   
 Monitoring the first of these two fields over almost
3 hours we could not detect any modulation in the optical
magnitude of any of these candidates. For a galaxy with a
reduced metallicity, like M31, we expect to find close binary SuSo
of the type described by van den Heuvel et al., 1992, with
orbital periods $\geq$12 hours. The $\simeq$4 hours orbital
period binaries among SuSo
are less likely to be members of M31, and our sampling
interval might have not been sufficient to detect modulations in the 0. While we plan monitoring the field over
 longer time
periods,  spectroscopic information would be very desirable.

\begin{figure}
%\vspace{6 true cm}
\hspace{3 true cm} \psfig{figure=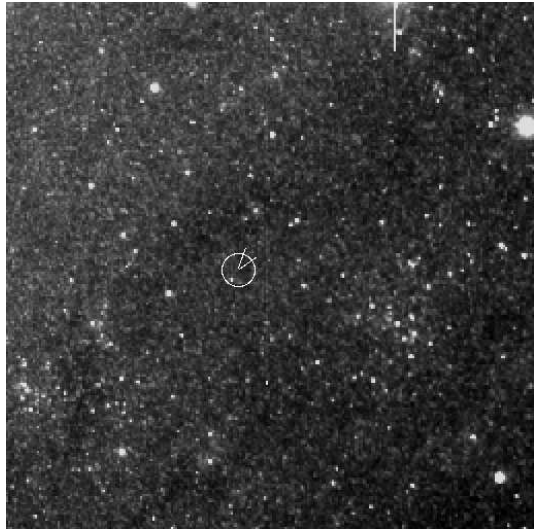,width=7cm}
\caption{The field of RXJ0044+4118 observed with
the R filter at the WIYN telescope in August 1999: 
 the arrow in the circle indicates the position of the 18th magnitude
star observed in September 1990 by Magnier et al.
(1992) and by Nedialkov and Tikhonov. The dimension of the image
is about 3.4 arcmin.}
\end{figure}
\section{An optical nova as possible counterpart}

In the field of RX J0044+4118, we have instead 
identified an optically  variable star, that we suggest
to be a classical or recurrent
nova, and also the optical counterpart of the 
supersoft X-ray source. In Fig. 1 we show the R image of
the field and indicate the position where  Magnier et al.
(1992) detected an object at  B=18.217 
with color index B-V=0.187,  at coordinates $\alpha$(2000)=00,44,04.71 and
$\delta$(2000)=41,18,21.6  (1.7 arcsec away from the ROSAT position listed
in the merged catalog in HEASARC), in observations
done between September 12 and 27 1990.
 A few days later the same object
was found to be at B=18.1 in a photographic plate taken at 1m telescope
at SAO RAN in Russia on September 21 1990. The object does  NOT appear 
in the blue POSS
(limiting magnitude V$\leq$22)  neither in plates taken in 
several previous and following years with the 2m telescope
at of the Bulgarian Academy of Science  at Rozhen in Bulgaria
and with the 1m telescope at SAO RAN, Russia, (limiting
magnitudes$\simeq$21), and finally not in WIYN images taken by us in August 1998
in U and R 
(the photometry was {\it complete} to a limit R$\leq$22,
but several objects were still detected at magnitudes R$\simeq$24). 
 We concluded that the  identified variable was most
likely a nova in M31, and that the  supersoft X-ray source,
detected in 1993, is associated with it. The position
of the supersoft X-ray source was still measured in ROSAT HRI observations
in 1996, so at that time the nova must have
still been ``on'' - i.e. the white
dwarf was burning hydrogen
in a shell. Given the optical magnitude observed in September 1990,
the nova outburst must have occurred during that
year and the hot remnant
appeared as a supersoft X-ray source for at least 6 years.  
Statistical considerations on the frequency and location
of novae in M31 suggest that the probability of a random
coincidence is very small (e.g. Capaccioli et al. 1989).

\section{Classical and recurrent novae as supersoft X-ray sources}

Up to now only 4 classical novae and one recurrent nova
in the Galaxy
have been identified with supersoft X-ray sources.
The existence of the hot white dwarf after the outburst has very
important consequences for the mechanisms and evolution
of nova systems (see accompaining poster by Orio \& Parmar).
Identifying classical or recurrent novae in M31 with 
SuSo allows much better statistics and gives us new insight  
into the secular evolution of nova systems.

\end{document}